\documentclass[12pt,halfline,a4paper]{ouparticle}

\usepackage[utf8]{inputenc} 
\usepackage[T1]{fontenc}    
\usepackage{hyperref}       
\usepackage{url}            
\usepackage{booktabs}       
\usepackage{float}
\usepackage{multirow}
\usepackage{multicol}
\usepackage{amsfonts}       
\usepackage{nicefrac}       
\usepackage{microtype}      
\usepackage{lipsum}
\usepackage{fancyhdr}       
\usepackage{amsfonts, amssymb, amsmath, amsthm, mathtools, bm}
\usepackage{bbm}
\usepackage{graphicx}       
\graphicspath{{media/}}     
\usepackage{derivative}
\usepackage[dvipsnames]{xcolor}
\usepackage{dsfont}

\usepackage[style=authoryear, backend=biber, sorting=nyt, maxbibnames=5, maxcitenames=2, natbib, uniquelist=false]{biblatex}
\bibliography{references.bib}

\begin{document}


\title{Flexible unimodal density estimation in hidden Markov models}

\author{Jan-Ole Koslik, Fanny Dupont, Marie Auger-Méthé, Marianne Marcoux, Nigel Hussey, Nancy Heckman}

\abstract{
\begin{enumerate}
    \item Hidden Markov models (HMMs) are powerful tools for modelling time-series data with underlying state structure. However, selecting appropriate parametric forms for the state-dependent 
    distributions is often challenging and can lead to model misspecification. To address this, P-spline-based nonparametric estimation of state-dependent densities has been proposed. While offering great flexibility, these approaches can result in overly complex densities (e.g.\ bimodal) 
    that hinder interpretability.
    \item We propose a straightforward method that builds on shape-constrained spline theory to enforce unimodality in the estimated state-dependent densities through enforcing unimodality of the spline coefficients. This constraint strikes a practical balance between model flexibility, interpretability, and parsimony.
    \item Through two simulation studies and a real-world case study using narwhal (\textit{Monodon monoceros}) dive data, we demonstrate the proposed approach yields more stable estimates compared to fully flexible, unconstrained models improving model performance and interpretability.
    \item Our method bridges a key methodological gap, by providing a parsimonious HMM framework that balances the interpretability of parametric models with the flexibility of nonparametric estimation. This provides ecologists with a powerful tool to derive ecologically meaningful inference from telemetry data while avoiding the pitfalls of overly complex models.
\end{enumerate}
}

\date{\today}

\keywords{animal behaviour, dive, hidden Markov model, non-parametric, P-spline, B-spline, unimodal, constrained density estimation}

\maketitle

\section{Introduction}

Hidden Markov models (HMMs) are a popular modelling framework to analyse time-series data derived from various fields, including speech recognition (\citealp{rabiner1989tutorial}), finance (\citealp{bhar2004hidden}), physics (\citealp{bechhoefer_hidden_2015}), biology (\citealp{sippl1999biological}) and ecology (\citealp{mcclintock2020uncovering}). Their popularity is a result of their ability to model an observed time series, and add others that are broder as arising from a sequence of discrete unobserved states that usually represent the process of interest, such as the purchase types of a consumer (\citealp{srivastava_credit_2008}), phonemes in speech (\citealp{rabiner1989tutorial}) or the behavioural state of an animal (\citealp{morales_extracting_2004}). Thus, HMMs are well suited to ecological studies, where data are inherently complex and critical information can be missing (\citealp{mcclintock2020uncovering}). They have been used to analyse animal movement data, providing novel insights into hunting strategies (\citealp{heerah_coupling_2017}), fine-scale habitat selection (\citealp{klappstein_flexible_2023}), and activity patterns (\citealp{mcrae2024killer}).

The success of HMMs in movement ecology has been facilitated by the development of accessible \texttt{R} packages such as \texttt{moveHMM} \citep{michelot2016movehmm}, \texttt{momentuHMM} \citep{mcclintock2018momentuhmm} and  \texttt{hmmTMB} (\citealp{michelot2022hmmtmb}), which enable researchers to fit complex models efficiently. However, these packages rely on parametric distributions to describe the observation process (e.g., gamma, normal, von Mises). Such assumptions are frequently unrealistic, as these models fail to account for the complexity of real data. This issue is compounded by the inherent challenge of selecting appropriate state-dependent distributions, as the latent nature of hidden states makes state-specific exploratory data analysis difficult. Consequently, state-dependent distributions are often misspecified and order selection methods are unreliable \citep{pohle2017selecting}. While recent advances have improved order selection methods for HMMs under misspecification, they do not fully resolve the problem of misspecified state-dependent distributions (\citealp{zou_order_2024}; \citealp{Dechaumaray2024}; \citealp{dupont2024improved}). 
Hence, it can be beneficial to drop the parametric assumption and use a more flexible class of density estimates in HMMs.

Most applications of nonparametric HMM estimation have relied on penalised B-splines, referred to as P-splines (\citealp{langrock2015nonparametric, langrock2017markov, langrock2018spline}). While other approaches, such as unpenalised B-splines \citep{titman2011flexible, chen2024bayesian} and K-nearest-neighbours methods \citep{mehrotra2005nonparametric} have been explored, P-splines are often used because they offer numerical stability and model flexibility. 
\citet{langrock2015nonparametric} highlighted the flexibility of nonparametric HMMs for estimating state-dependent distributions of Blainville's beaked whale (\textit{Mesoplodon densirostris}) dive data using P-splines.
When applied to beaked whale dive data, this approach effectively identified a parsimonious set of states that were biologically meaningful, outperforming traditional parametric methods. However, this approach presents three key challenges: (1) computational intensity, (2) technical implementation barriers that limit accessibility for ecologists, and (3) potential excessive model flexibility that can lead to overfitting. Specifically, there can be cases of multimodality that produce overfitting, as the presence of multiple modes can lead the model to fit noise or spurious patterns in the data rather than accurately represent the underlying process.
Such multimodality may obscure biological interpretation, as behavioural states (e.g., foraging vs. travelling) are most meaningfully and predominantly represented by unimodal distributions that clearly distinguish distinct movement patterns (\citealp{ngo2019understanding}; \citealp{mcclintock2020uncovering}).

We overcome these limitations by combining recent advances in smooth density estimation for HMMs (\citealp{koslik2024efficient}) with unimodality constraints (\citealp{kollmann2014unimodal}; \citealp{pya2015shape}). These constraints guarantee realistic unimodal state-dependent distributions while preserving model flexibility, yielding clearer interpretation of biological patterns from non-parametric fits. Our method is implemented in the \texttt{R} package \texttt{LaMa} (\citealp{koslikLaMa2025}), provide ecologists with an efficient and practical tool for reliable biologging data analysis. We begin in Section \ref{sec:methods} by describing HMMs, as well as nonparametric estimation of state-dependent densities with P-splines. We also show how unimodality constraints can be imposed on the densities. In Section \ref{sec:simulations}, we demonstrate the efficacy of the approach through a simulation study. We compare our framework with parametric and nonparametric unconstrained methods, and show how the proposed method is more robust to misspecification of the state process model. Lastly, Section \ref{sec:case_study} presents a real-world application to examine narwhal (\textit{Monodon monoceros}) time series dive data.

\section{Methods}
\label{sec:methods}

\subsection{Hidden Markov models}

A standard hidden Markov model comprises two stochastic processes: an observed process $\{X_t\}$ and a hidden process $\{S_t\}$. The unobserved process is an $N$-state Markov chain that satisfies the Markov property $$\Pr({S}_{t+1} = {s}_{t+1}|{S}_t = {s}_t,\ldots,{S}_1 = {s}_1) = \Pr({S}_{t+1 }= {s}_{t+1}|{S}_t = {s}_t),$$
where each $S_t$ only take values in $\{1, \ldots, N\}$.  
In movement ecology, the states are typically interpreted as a proxy for behaviours, such as travelling or foraging (e.g., \citealp{patterson2017statistical}, \citealp{leos2017analysis}). The distribution of $X_t$, given the corresponding hidden state $S_t$, is independent of all other observations and hidden states. If $S_t=i$, then we denote the state-dependent distribution of $X_t$ by $p_i(\cdot) = p(\cdot \mid S_t = i)$. Due to its Markovian nature, the state process is fully specified by its initial distribution $\bm{\delta} = \bigl(\Pr(S_1 = 1), \dotsc, \Pr(S_1 = N) \bigr)$ and one-step transition probabilities $\gamma_{ij} = \Pr(S_{t+1} = j \mid S_{t} = i)$ which are commonly summarised in a transition probability matrix (t.p.m.) $\bm{\Gamma} = (\gamma_{ij})_{i,j = 1, \dotsc, N}$. In principle, this Markov chain can also be time-inhomogeneous (e.g.,\ if the transition probabilities depend on covariates), but for simplicity, we focus on time-homogeneous Markov chains.


While traditionally, the expectation maximisation (EM) algorithm has been used for HMM inference \citep{baum1970maximization,leroux1992maximum}, we focus on direct numerical maximum likelihood estimation, due to it being the more user-friendly alternative \citep{macdonald2014numerical}.
Exploiting the Markov property enables a fast recursive scheme to evaluate the likelihood using the forward algorithm \citep{zucchini2016hidden, mews2025build}, leading to the closed-form expression for the HMM likelihood
\begin{equation}
\label{eqn:likelihood}
\mathcal{L}(\bm{\theta}) = \bm{\delta} \bm{P}(x_1) \bm{\Gamma}\bm{P}(x_2) \dotsc \bm{\Gamma} \bm{P}(x_T) \bm{1}^\intercal,
\end{equation}
where $\bm{P}(x_t) = \text{diag}\bigl(p_1(x_t), \dotsc, p_N(x_t)\bigr)$ and $\bm{1}$ is a row vector of ones. Implementing the recursive scheme to calculate the logarithm of Eq \eqref{eqn:likelihood} in a numerically stable way is straightforward \citep{lystig2002exact}, and the specific form of $\bm{P}(x_t)$ is highly modular, which we exploit to incorporate flexible state-dependent densities.


\subsection{Flexible density estimation using P-splines}




Following \citet{langrock2015nonparametric}, we model the state-dependent distributions with a linear combination of B-splines and, in fitting, penalise the log-likelihood to prevent overfitting.
Generally, 
the state-dependent densities 
can be expressed as finite linear combinations of predefined fixed basis functions $\phi_1, \dotsc, \phi_K$, taking the form
\begin{equation}
\label{eqn:splinedens}
    p_i(x) = \sum_{k=1}^K \alpha_k^{(i)} \phi_k(x), \quad i = 1, \dotsc, N,
\end{equation}
where the flexibility of $p_i$ depends on $K$ and the $\phi_k$'s. 
For $p_i$ to be a valid probability density function, we take the basis functions to be densities (i.e.\ they are non-negative and integrate to one) and the coefficients $\alpha^{(i)}_{k}$ to satisfy $\sum_{k=1}^K \alpha_k^{(i)} = 1$, with $\alpha_k^{(i)} \geq 0$ for all $k = 1, \dotsc, K$.
The constraints on the $\alpha^{(i)}_k$'s can be enforced by expressing the coefficients as transformations of unconstrained parameters $\beta_k^{(i)} \in \mathbb{R}$ using the inverse multinomial logistic link function: $\alpha_k^{(i)} = \exp(\beta_k^{(i)})/\sum_{j=1}^K \exp(\beta_j^{(i)})$. For identifiability, we set $\beta_K^{(i)}$ to zero.

\begin{figure}
    \centering
    \includegraphics[width=1\linewidth]{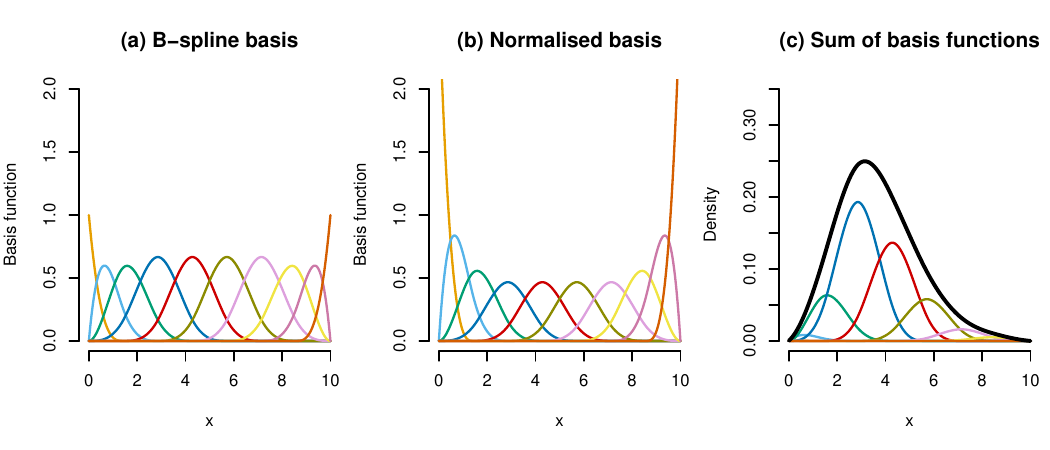}
    \caption{$K=10$ cubic B-spline basis functions (a), normalised version of these basis functions such that each one integrates to 1 (b), and the same normalised basis functions, weighted by nonnegative weights that sum to 1, with their sum giving rise to a right-skewed density (thick black line) (c).} 
    \label{fig:B_splines}
\end{figure}

While any probability density functions can be used for the basis expansion defining $p_i$,
for the real-valued $X_t$ we consider here, normalised B-spline basis functions similar to \citet{schellhase2012density} and \citet{langrock2015nonparametric} are a particularly convenient and numerically stable choice due to their local support.
We set
\begin{equation}
\label{eqn:normalize}
\phi_k(x) = \frac{B_k(x) }{\int B_k(x) \; dx}
\equiv \nu_k B_k(x), \quad k=1,\ldots, K,
\end{equation}
where $\{B_1,\ldots,B_K\}$ is the usual cubic B-spline basis with specified knots (see \cite{eilers.marx.1996}, or \cite{deboor}).
The resulting $p_i$ is a cubic spline, that is, $p_i$ is a piecewise cubic polynomial, 
where the pieces are defined in terms of predefined knots
and are joined to ensure continuity of the second derivative. 
See Figure~\ref{fig:B_splines} for visual intuition on this type of basis and how it can give rise to very flexible densities.
The number of basis functions $K$ should be sufficiently large to allow any flexible shape that could plausibly explain the observed data. However, excessive flexibility without proper constraints can lead to overfitting. 

To prevent overfitting, we constrain the flexibility of the density functions by adding to the log likelihood function a penalty on each $p_i$'s curvature.
Specifically, we penalise $p_i$ by penalising the vector of second differences of the coefficient vector $\bm{\beta}^{(i)} = (\beta_1^{(i)}, \dotsc, \beta_K^{(i)})^\intercal$. 
The second differences vector of a $K$-vector $\bm{\beta}$, denoted $\bm{\Delta}_2 \bm{\beta}$,
has the $k$th component equal to $(\beta_{k+2} - \beta_{k+1}) - (\beta_{k+1} - \beta_k) = \beta_{k+2} - 2 \beta_{k+1} + \beta_k$. 
The penalty on $p_i$ is
\begin{equation}
\label{eqn: penalty}
-\lambda_i || \bm{\Delta}_2 \bm{\beta}^{(i)}||^2 
\equiv  -\lambda_i \bm{\beta}^{(i)\intercal} \bm{S} \bm{\beta}^{(i)},
\end{equation}
a quadratic form in terms of  the $K \times K$ matrix  $\bm{S}$.
The combination of using this very specific penalty on the coefficient vector only and a B-spline basis was introduced by \citet{eilers.marx.1996} to shorten computation time, and was named \textit{P-spline}.
This computationally convenient regularisation ensures that, while the model remains flexible to capture true patterns, it avoids overfitting through excessive oscillations in the estimated density.

As discussed by \citet{lang2004bayesian} and \citet{koslik2024efficient}, adding the sum of these $N$ quadratic forms to the log-likelihood is equivalent to using an improper multivariate Gaussian prior on each $\bm{\beta}^{(i)}$, specifically $\bm{\beta}^{(i)} \sim \mathcal{N}(\bm{0}, \bm{S}^- / \lambda_i)$, where $\bm{S}^-$ is the generalised inverse of $\bm{S}$, assuming a priori independence. This random effects representation enables efficient estimation of both the $\lambda_i$'s and $\bm{\beta}^{(i)}$'s via restricted maximum likelihood (REML). The resulting estimation procedure, as described by \citet{koslik2024efficient}, uses an iterative algorithm that  (approximately) estimates the smoothing parameters in a data-driven way and typically converges within a moderate number of penalised model fits.

\subsection{Unimodality constraints}

\label{subsec:unimodality}

The flexibility provided by P-spline-based density estimation can be problematic in many contexts; thus we propose to reduce flexibility by limiting each emission distribution to be unimodal. For a spline function to be unimodal, it suffices that the coefficients in its B-spline basis expansion form a unimodal sequence \citep{schoenberg1967on, goodman1985properties, kollmann2014unimodal}. Noting that the expansion in Equation \eqref{eqn:splinedens} is in terms of a normalised B-spline basis instead of a regular one, to ensure unimodality, we use Equation \eqref{eqn:normalize} to rewrite \eqref{eqn:splinedens} as
$$
p_i(x) = \sum_{k=1}^K \alpha_k^{(i)} \nu_k B_k(x)
\equiv \sum_{k=1}^K \Tilde{\alpha}_k^{(i)} B_k(x).
$$
 If the $\Tilde{\alpha}^{(i)}_k$s form a unimodal sequence, then $p_i$ is unimodal.  Therefore, we specify that, for some $m=1, \dotsc, K$, 
$$
\Tilde{\alpha}^{(1)},\dotsc,\Tilde{\alpha}^{(m)}
{\rm~is~a~nondecreasing~sequence}
$$
and
$$
\Tilde{\alpha}^{(m)},\dotsc,\Tilde{\alpha}^{(K)}
{\rm~is~a~nonincreasing~sequence}.
$$

Letting $\Tilde{\bm{\alpha}}^{(i)}$ be the $K$-vector with $k$th component  $\Tilde{\alpha}_k^{(i)}$, we can can write these inequality constraints as 
$$
 \bm{C}_m \Tilde{\bm{\alpha}}^{(i)} \geq 0
$$
where, for $m=2,\dotsc K-1$, 
\[
\bm{C}_m = \left(\begin{array}{ccccccc}
    -1 & 1 &  & & \cdots & & 0 \\
     & \ddots & \ddots & & & & \\
     & & -1 & 1 & & &  \vdots\\
     \vdots & & & 1 & -1 & & \\ 
     & & & & \ddots & \ddots & \\
     0 & & & & & 1 & -1 \\
\end{array}\right) \begin{array}{l}
    \\
    \\
    \\
    \leftarrow m\text{-th row} \\
    \\
    \\
\end{array}
\]
with the sign change occurring in the $m$-th row. For $m=1$, the matrix $C_1$ has no sign change, but rather has all rows containing 1 followed by -1, yielding a monotone decreasing density. For $m=K$, the matrix $\bm{C}_K$ also has no sign change, but rather has all rows containing -1 followed by 1, yielding a monotone increasing density.
The index $m$ defining the sign change in the coefficient sequence has substantial influence on the position of the mode of $p_i$ but does not determine it uniquely. Thus, from now on, we will carefully use language to distinguish the coefficient-mode $m$ and the mode of $p_i$.

To avoid computations that restrict the $\Tilde{\alpha}_k^{(i)}$s to be nonnegative and sum to 1, we use the multinomial logistic link and define
$$
\Tilde{\beta}_k^{(i)} \equiv \log (\Tilde{\alpha}_k^{(i)}/ \Tilde{\alpha}_K^{(i)}).
$$ 
Thus, the unimodality constraint $\bm{C}_m \Tilde{\bm{\alpha}}^{(i)} \ge 0$ can be written as
\begin{equation}
    \label{eqn:constraint}
    \bm{C}_m \Tilde{\bm{\beta}}^{(i)} \geq 0.
\end{equation}
To use the penalty on the original ${\bm{\beta}}^{(i)}$'s in (\ref{eqn: penalty}), we see that
$\Tilde{\beta}_k^{(i)} 
= \log (\alpha_k^{(i)}/ \alpha_K^{(i)}) + \log (\nu_k / \nu_K) = \beta_k^{(i)} + \log (\nu_k / \nu_K)$,
which is just the original coefficient with a fixed term added that corrects for the basis function rescaling.
The constraint \eqref{eqn:constraint} can be viewed as a truncation of the multivariate Gaussian prior distribution on $\bm{\beta}^{(i)}$, implied by the penalty in \eqref{eqn: penalty}. The truncation is to the subset of $\mathbb{R}^K$, 
$\mathcal{S}_{m} = \{ \bm{\beta}^{(i)}: \bm{C}_{m} \Tilde{\bm{\beta}}^{(i)} \geq 0\}$.

For better computational performance, we turn away from constrained optimisation. We replace the hard constraint in \eqref{eqn:constraint} with a penalty that approximates the log-density of the truncated Gaussian by giving large negative penalties on the truncation set.
Therefore, we consider a penalty of the form
$$
P_i(m_i) = \sum_{k=1}^K - \text{min}\bigl((\bm{C}_{m_i} \Tilde{\bm{\beta}}^{(i)})_k, 0\bigr)
$$
for each state $i = 1, \dotsc, N$.
To make the above expression differentiable, we use the smooth approximation
$$
- \text{min}(x, 0) \approx \; \frac{1}{\rho} \log\bigl[1 + \exp(- \rho x)\bigr], 
$$
where the hyperparameter $\rho$ controls the smoothness of the approximation. As $\rho \to \infty$ 
the approximation becomes more accurate.
We denote this approximation to $P_i$ as $\Tilde{P}_i$. Thus, we maximise the doubly penalised log likelihood, with likelihood $\mathcal{L}$ as in (\ref{eqn:likelihood}) and with penalties (\ref{eqn: penalty}) and $\Tilde{P}_i(m_i)$  with $\rho = 20$. That is, we maximise
\[
\log \mathcal{L}(\bm{\theta}) - \sum_{i=1}^N \lambda_i \bm{\beta}^{(i) \intercal}\bm{S} \bm{\beta}^{(i)}
- \kappa \sum_{i = 1}^N \Tilde{P}_i(m_i), 
\]
over the collection of all model parameters $\bm{\theta}$ --- including the $\bm{\beta}^{(i)}$'s, and $m_1,\ldots, m_N$. The constant $\kappa$ determines the ``sharpness'' of the unimodality constraint.
As a consequence of using penalised instead of constrained optimisation, for small $\kappa$ values, the estimated state-dependent densities may be multimodal but will be forced to unimodality for a sufficiently large $\kappa$. 
Furthermore, by using the smooth approximation 
of the minimum, we in fact enforce \textit{strict} unimodality in the coefficient sequence, which results in unique estimates of $m_1,\ldots, m_N$ and in the estimated densities tending to avoid ``flat tops''. We found that, in practice, our approach of approximating the optimisation problem shortened computation time without negatively impacting estimation (see Section~\ref{subsec:simple_sim}).

In non-HMM settings, other methods have been developed  to estimate a unimodal density function. See, for instance, \citet{meyer2012}.  While placing a shape constraint on a regression function is relatively easy (see, e.g., \cite{ramsay1988monotone}, 
\cite{meyer2008regression}, and \cite{pya2015shape}) placing a constraint on a density is computationally more challenging, due to the requirement that the density integrate to 1. Our approach, in the complex HMM setting, is intuitive and computationally stable.

\subsection{Practical implementation}

\label{subsec:practical}


The described methods are implemented in the \texttt{R} package \texttt{LaMa} \citep{koslikLaMa2025}. To enable convenient but highly customisable B-spline-based density estimation, the package contains helper functions to set the basis and penalty but the user is expected to write a custom penalised likelihood function.

Specifically, we provide the function \texttt{smooth\_dens\_construct()} for easy setup of model matrices as well as the function \texttt{penalty\_uni()} to compute the unimodality penalty.
Once the doubly penalised likelihood function --- including a smoothness penalty and the unimodality penalty --- is defined, it can be passed to \texttt{qreml()}, which performs the automatic smoothness selection procedure as outlined in \citet{koslik2024efficient}.
In the online Supplementary Materials, we provide an in-depth tutorial similar to the case study, demonstrating how models with B-spline based densities with and without unimodality penalties can be both implemented using \texttt{LaMa}.


\subsubsection*{Mode finding}

In general, to maximise the doubly penalised log likelihood, one needs to search the full ordered grid
$$
1 \leq m_1 \le m_2 \le \dotsc \leq m_N \leq K
$$
for the coefficient modes in all $N$ states. Searching this grid over all possible combinations of mode positions is computationally expensive. Hence, we propose a more practical approach.
We first fit a nonparametric model without the unimodality constraint. Such a model then yields one or more modes in the coefficient sequence for each state. If for state $i$, it is clear that all but one mode is spurious, we take the value $m_i$ and explore the two indices before and after it as well. If state $i$ yields several competing modes of similar magnitude, we explore these as well as all potential positions in between.
If in the nonparametric model fit, a  state-dependent distribution is monotonically decreasing, we fix the mode index to the value found in the nonparametric fit.
This approach thus gives rise to typically much smaller grids to search, making it more convenient.

\subsubsection*{Initial parameter values for state-dependent densities}

Robust initialisation of the coefficients of the B-spline-based densities can be challenging. 
We choose a parametric ``target'' density $p_i^*$ and then initialise with a B-spline-based density $p_i$  that is close to $p_i^*$. 
While there are many ways of choosing $p_i$ close to a specified $p_i^*$,  we use the following heuristic approach, which is very fast and showed generally good performance. 
The idea is that a basis function's contribution to $p_i$ should be large if the values of $p_i^*$ are large on the basis function's support.
To implement this, we first choose $K$ points, $x_1,\ldots, x_K$, with $x_k$ a ``typical'' point in the support of the normalized basis function $\phi_k$ in (\ref{eqn:normalize}). Specifically, we set $x_k$ to the expected value, computed via the density $\phi_k$, that is we take $x_k = \int x \phi_k(x)~dx$.  If the value $p_i^*(x_k)$ is large, then we will take the coefficient of $\phi_k$ in (\ref{eqn:splinedens}) to be large.  
To avoid constraints on the $\alpha_i^{(k)}$'s, we reparameterise via the multinomial link function, and set the $k$th component of $\bm{\beta}^{(i)}$ equal to the logarithm of $p_i^*(x_k)$. Thus a large value of $p_i^*(x_k)$ will result in a large value of the $k$th component of $\bm{\beta}^{(i)}$.

For the parametric densities, the $p_i^*$s, we use normal densities for data on the real number line (as in the simulation experiments) and gamma densities for strictly positive data (as in the case study).
Hence, only means and standard deviations need to be specified (which are transformed to shape and scale for the gamma density).
This procedure is automated in \texttt{smooth\_dens\_construct()}, easily  allowing the quick calculation of initial coefficient values for user-specified state-dependent means and standard deviations.  We note that the target densities are only used at the initialisation stage.

\section{Simulation experiments}
\label{sec:simulations}

The following two simulation experiments demonstrate our approach's practical performance and robustness to misspecification. All code for reproducing the simulation experiments can be found 
in the online Supplementary Material.

\subsection{Validation}

\label{subsec:simple_sim}

In the first simulation experiment, we generate data from an HMM with unimodal parametric state-dependent distributions and compare the fit of four models.  The first two models compared are parametric: the true parametric model, and a misspecified one that uses a distribution commonly available in \texttt{R} packages used to fit HMMs. The second two models are nonparametric fits: one with no constraints on the state-dependent distributions and one that constrains the state-dependent distributions to be unimodal. We demonstrate that the constrained nonparametric fit performs well, even compared to fitting the true parametric model.

Specifically, we simulate a 2-state HMM, in which observations are generated from a \textit{skew normal} distribution with parameters $\xi = 0$, $\omega = 1$, $\alpha = 6$ in state~1 and a $t$-distribution with parameters $\mu = 3$, $\sigma = 1$, $\nu = 3$ in state~2.
The skew normal distribution is strongly right-skewed, while the $t$-distribution exhibits high kurtosis. Although both distributions are unimodal, they deviate substantially from the commonly used normal distribution and are not supported by widely used HMM software packages, such as \texttt{moveHMM} \citep{michelot2016movehmm},  \texttt{momentuHMM} \citep{mcclintock2018momentuhmm} or \texttt{hmmTMB} (\texttt{hmmTMB} includes the $t$-distribution but not the skew normal; \citealp{michelot2022hmmtmb}). 
Even if both distributions were available, identifying the correct parametric model formulation via model selection would be tedious and arriving at the appropriate specification unlikely. This highlights the value of nonparametric methods, which circumvent the need to pre-specify a suitable family of emission distributions. State sequences are generated by a Markov chain with transition probability matrix
$$\bm{\Gamma} = \begin{pmatrix}
    0.9 & 0.1 \\
    0.1 & 0.9
\end{pmatrix}$$
and initial distribution $\bm{\delta} = (0.5, 0.5)$. We simulated a total of 100 data sets, each comprising $T = 500$ observations.

\begin{figure}
    \centering
    \includegraphics[width=1\linewidth]{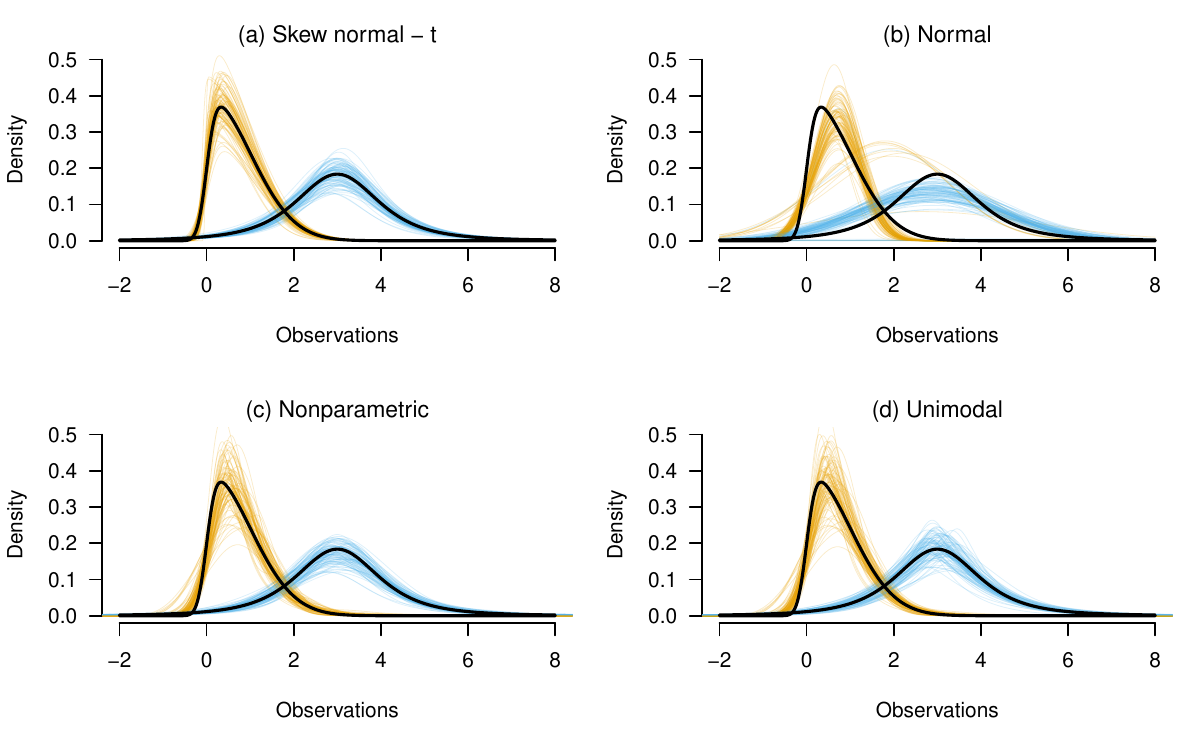}
    \caption{Estimated weighted state-dependent distributions from fitting the four different models, colour coded orange (state~1) and blue (state~2). Simulated data are generated from parametric state distributions skew normal and $t$. True densities are shown as thick black lines.}
    \label{fig:simple_sim}
\end{figure}

To each of the $100$ simulated tracks, we fitted the true model using skew normal and $t$ state-dependent distributions and a model with normal state-dependent distributions. 
When fitting the true model, we initalised the numerical estimation with the true parameter values. When fitting the incorrect normal model, we initialised with the mean and standard deviations of the generating skew normal and $t$-distributions.
For both models with nonparametric state-dependent distributions, we used $K = 40$ cubic B-spline basis functions for each state, with knots placed equidistantly in the range of the observations. 
To avoid convergence to a local optimum, for each simulated data set, we applied the initialisation strategy detailed in Section~\ref{subsec:practical} to a set of 15 initial means and standard deviations, drawn randomly around the data-generating means and standard deviations. 
We then picked the model that yielded the highest log-likelihood value after optimisation.



As expected, the correct parametric and both nonparametric models capture the shape of the true emission distribution (Figure~\ref{fig:simple_sim}). 
In contrast, the normal model performs poorly. Because it cannot account for the skewness in state~1 and kurtosis in state~2, the normal model results in large biases and in several cases converges to degenerate solutions that comprise effectively only one state.
The unimodal fits tend to be slightly more peaked around the mode compared to the unconstrained fits. 
This phenomenon can be explained by the interplay between approximating the constrained optimisation problem by a penalised optimisation problem and using a smooth approximation of the minimum function, leading to the penalty enforcing \textit{strict} monotonicity on both sides of the mode, hence avoiding \textit{flat tops} as explained in Section~\ref{subsec:unimodality}.

\begin{figure}
    \centering
    \includegraphics[width=0.65\linewidth]{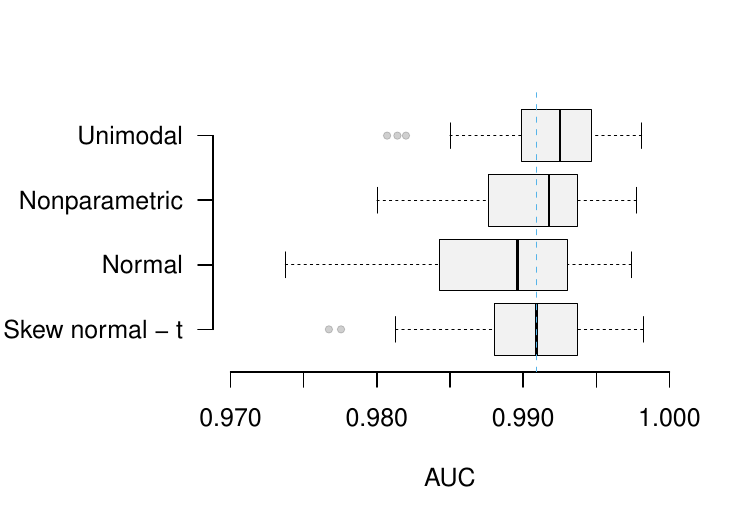}
    \caption{Boxplots of area under ROC curve (AUC) from a state-decoding using fits from the simulated data for the four different models. The blue dashed line shows the median AUC achieved by the correct model to aid visual comparison. A high value of AUC corresponds to better state identification.}
    \label{fig:simple_sim_AUC}
\end{figure}

To complement this qualitative assessment, we also consider a performance metric based on the goodness of state decoding. We work with locally decoded state probabilities
\[
p_t(j) = \Pr(S_t = j \mid X_1, \dotsc, X_T), \quad j = 1,2.
\]
For a particular classification threshold $\tau \in (0,1)$, if $p_t(1) > \tau$, the state at time $t$ is decoded as 1, otherwise 2. This gives rise to a decoded state sequence which can be compared to the true states simulated in that run. Varying the threshold from 0 to 1, we compute a receiver-operating characteristic (ROC) curve \citep{zweig1993receiver} for which we compute the area under the curve (AUC) \citep{hanley1982meaning} as a meaningful summary statistic for each model and simulation replicate. For AUC, a score of 1 corresponds to a perfect classifier and 0.5 to a fully random classifier.

The correctly specified parametric model and both nonparametric approaches perform well, each achieving a median AUC above 0.99 (Figure~\ref{fig:simple_sim_AUC}). 
In contrast, the misspecified normal model performs worse, with a median AUC slightly below 0.99 and a substantially longer left tail, indicating poorer performance in several replicates.
The minimally higher average performance of both nonparametric models over the true parametric model is likely due to sampling variability. 
The constrained model seems to perform slightly better than the unconstrained model, in terms of the AUC-distribution having a shorter right tail, but this difference might also be attributed to sampling variability.



\subsection{Robustness to state-process misspecification}

We now illustrate how constraining the state-dependent distributions to be unimodal can enhance the robustness of HMMs to misspecification in the state process. Such misspecification is common in practice because many real-world processes are only approximately Markovian and selecting an appropriate model for the state process is inherently challenging due to the process's latent nature.

To this end, we simulated data from a two-state hidden semi-Markov model (HSMM) \citep[][Chapter 12]{zucchini2016hidden} with unimodal emission distributions in each state. Hidden semi-Markov models generalise HMMs by allowing the hidden process to be a semi-Markov chain \citep{ferguson1980variable, guedon2003estimating}. For such a semi-Markov chain, the dwell-time distribution, i.e.\ the distribution of time-spent in a state, can be defined flexibly as opposed to the implied geometric distribution of a Markov chain.
Here, the dwell-time distribution in the first state is non-standard, while the second state follows a geometric dwell-time distribution associated to a conventional Markovian state. 
Specifically, the dwell times in state 1 were generated from a mixture of two Poisson distributions --- a Poisson(0.1) and a Poisson(15) --- with mixture weights 0.7 and 0.3, resulting in a bimodal probability mass function (Figure~\ref{fig:hsmm_sim_dwelltime} in the Supplementary Material).
This results in predominantly long runs of state 1 with occasional very short visits of one or two time steps, sandwiched between visits to state 2. In contrast, the dwell time in state 2 follows a geometric distribution with success probability $p = 1/10$, corresponding to an average dwell time of 10 observations and a self-transition probability of 0.9.
Conditional on the latent state sequence, observations were generated from gamma distributions with means 1 and 15 and standard deviations 1 and 4 for states 1 and 2, respectively. We simulated a total of 100 data sets of length 1000 from this model.

\begin{figure}
    \centering
    \includegraphics[width=1\linewidth]{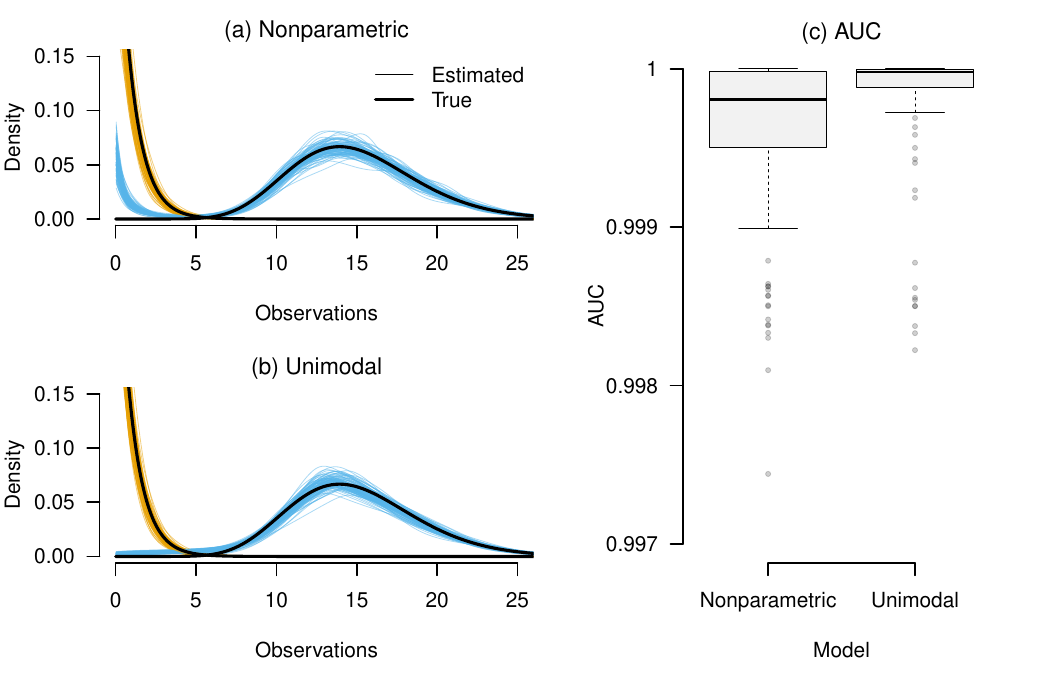}
    \caption{Estimated weighted state-dependent distributions for state 1 (orange) and state 2 (light blue) obtained from the parametric (a) and unimodality-constrained (b) nonparametric HMM as well as AUC values of both models (c). The true densities are shown as thick black lines.}
    \label{fig:sim_hsmm}
\end{figure}

To each of these data sets, we fitted an unconstrained nonparametric HMM with $K= 30$ B-spline basis functions for each state-dependent density, and the same model with a unimodality constraint added. For both models, we used 10 sets of random initial parameter values for each simulation replicate to avoid local maxima of the likelihood. 
We did not fit HSMMs here, as this scenario only serves as an illustrative example of state-process misspecification, while the specific reasons for misspecification are usually unknown in practice.
HSMMs come with their own challenges; for example, including covariates in the state process has only been made available recently \citep{lagona2025nonhomogeneous, koslik2025hidden} and the combination of semi-Markovian state processes and nonparametric emission densities has not been studied.

While the unconstrained nonparametric model yielded a bimodal state-dependent distribution for state~2 in 100\% of the cases, this never happened with the constrained model (Figure~\ref{fig:sim_hsmm}).
Due to this bias, the persistence in state 2 was also severely overestimated by the unconstrained models, with an average of $\Hat{\gamma}_{22} = 0.966$ (with standard deviation 0.007), which results in an average state dwell time of 31.5 (with standard deviation 7.1), with the true value being 10.
While in the constrained model the state process was also misspecified, the unimodality constraint meant that the fit 
correctly allocated small observations to state~1. Thus, with the unimodal model, the overestimation of persistence in state~2 was less severe, with an average of $\Hat{\gamma}_{22} = 0.924$ (with standard deviation 0.022), resulting in an average state dwell time of 15.4 (with standard deviation 4.8).
We also computed AUC to assess decoding accuracy across all simulations. Due to the clear state separation, both models perform well according to AUC. However, the constrained model consistently achieves higher decoding accuracy, demonstrating the benefits of robust emission modelling (Figure~\ref{fig:sim_hsmm}).

The relatively poor performance of the unconstrained model arises from the combination of highly flexible nonparametric estimation of the emission distributions and misspecification of the state process.
In this setup, short visits to state~1  --- producing small observations amid long sequences of state 2 --- tend to be misinterpreted as arising from state~2 itself.
This is because very short stays are unlikely under the Markov assumption, hence the model tends to compensate by assigning a second mode to the emission distribution of state~2, thereby introducing artificial bimodality.
By enforcing unimodality on the emission distributions, the constrained model is more robust to this type of misspecification. It reflects prior knowledge that bimodal emission distributions are not expected and thereby avoids ``misclassifying'' small observations from short state~1 episodes as originating from state~2 during the estimation process. 




\section{Case study: Narwhal dive data}
\label{sec:case_study}

We illustrate our method using dive data derived from narwhal for a four month period (August - November 2017) tagged in the Qikiqtaaluk (Baffin) region in Nunavut, Canada. This individual (ID 17001) was equipped with a TDR10 satellite transmitting tag (Wildlife Computers Inc$.$) in Tremblay Sound \citep[72°30'N, 80°45'W; permit AUP 40, S-17/18-1017-NU; capture and tagging protocols are detailed in][]{shuert2022decadal}. All capture and tagging protocols were approved by the Fisheries and Oceans Animal Care Committee and a Licence for Scientific Purposes was granted. While the satellite telemetry device provided several data streams, we focused exclusively on the depth profile recorded at 75-second intervals by the time-depth recorder. To account for potential short-term behavioural changes following capture, we excluded the first 24 hours of post-tagging data \citep{shuert2021assessing, shuert2022decadal}. Maximum dive depth is a key metric to discriminate between distinct diving behaviours for whales that are specialized in deep foraging (\citealp{deruiter2017multivariate}; \citealp{shuert2025putting}).
Thus, we used the \texttt{R} package \texttt{diveMove} (\citealp{luque2007introduction}) to extract a time series of the maximum depth (in meters) of subsequent dives from the raw depth data. As in \citet{ngo2019understanding}, dives were defined as recorded depth below $20$ meters. Thus, a dive corresponds to any descent from the surface to a depth greater than 20 meters, followed by the ascent back to the surface. This preprocessing yielded a total of $3114$ dives with values of maximum depth ranging between 20 and 979 meters. To facilitate the analysis, we subtracted 20 meters from each observation, thereby obtaining data supported on the positive real number line. The aim is to use HMMs to classify dives into what we will call: \textit{shallow}, \textit{mid}, and \textit{deep} diving behaviours. Previous studies have interpreted \textit{shallow} dives as potentially associated with transit, social activities, and resting; \textit{mid} dives and \textit{deep} dives with shallow and deep foraging-like behaviours, respectively (\citealp{ngo2019understanding}; \citealp{shuert2025putting}).


We fitted three candidate models: a parametric 3-state gamma HMM as used in \citet{ngo2019understanding} and \citet{shuert2022decadal} to model narwhal dive data, a nonparametric 3-state model with P-spline-based state-dependent densities, and the same model with an additional unimodality constraint, approximated by a penalty function as described in Section \ref{subsec:unimodality}.
For the parametric model, we initialised the maximum likelihood estimation with 500 random initial values to help avoid convergence to a local maximum of the likelihood.
For both nonparametric models, we used $K = 30$ normalised B-spline basis functions which we deem to be a sufficient number to cover all possible shapes of the emission distributions. Knot locations were chosen equidistantly in the range of the data. 
Using the qREML procedure, the model was fitted with 200 sets of random initial values, including random starting values for the smoothness parameters.
For the constrained model, we did not search the entire ordered grid for mode positions of the spline coefficients. Instead, for the shallow diving state, we fixed the coefficient mode at one because both previous fitted models showed a monotonically decreasing density in the first state. 
For the mid and deep states, we used the largest mode of the estimated coefficients from the unconstrained nonparametric fit as a baseline and explored them as well as the four neighbouring indices. We did not explore other modes than the largest one. This means that 25 models needed to be fitted, each with similar computational complexity as the nonparametric model, which is however quite modest here (estimating one model took about 30 seconds on an Apple M2 chip).

\begin{figure}
    \centering
    \includegraphics[width=1\linewidth]{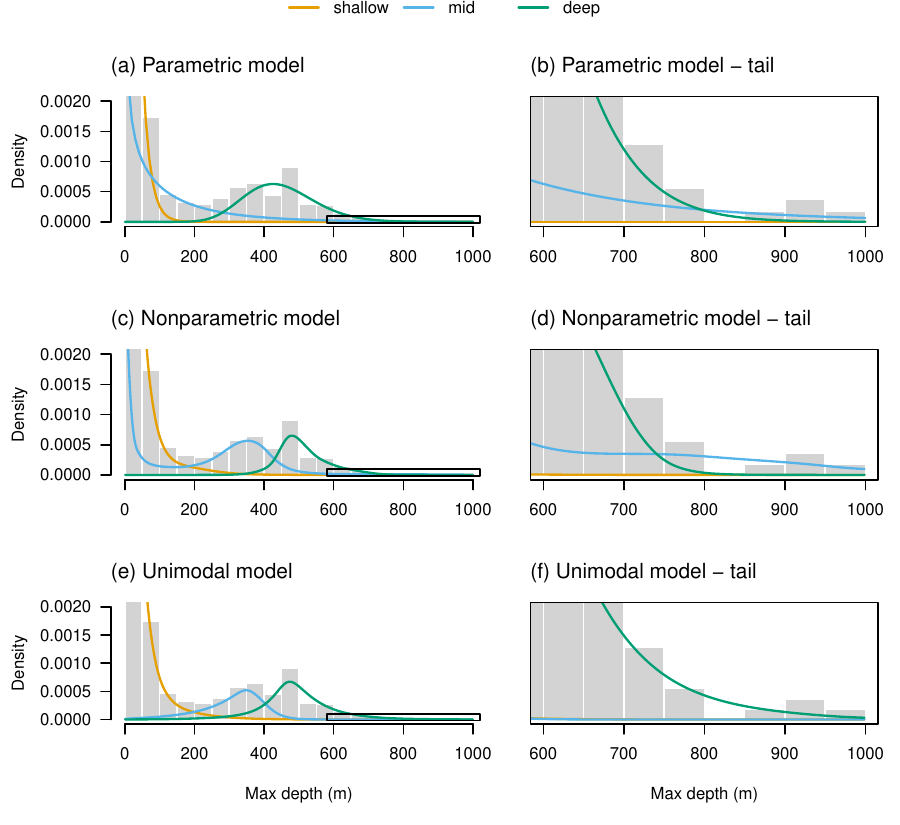}
    \caption{Estimated (weighted) state-dependent distributions of the three models fitted to the narwhal data, complemented with the histogram. Panels (b), (d), and (f) show a magnified version of the distributions, focusing on the tail (indicated by the black box in panels (a), (c), and (e)). The first bar in the full histograms is truncated at the top for better readability.}
    \label{fig:narwhal_models}
\end{figure}

While data appear to be a mixture of \textit{shallow} (<150 meters), \textit{mid} (150-500 meters), and \textit{deep} (> 500 meters) dives, the gamma distribution is not flexible enough to yield state-dependent distributions that distinctly separate dives into these states (Figure~\ref{fig:narwhal_models} a).
The whale's state-switching behaviour forces the model to use state~2 for both very small observations and those around 400 meters. The heavy right tail associated with state~2 absorbs some of the observations around 800 meters. Consequently, the mode corresponding to deep dives (state~3) is poorly captured. 
The overlap between the state-dependent distributions makes the ecological interpretation of the three states challenging. The labels \textit{shallow}, \textit{mid}, and \textit{deep} do not match the fitted model.

\begin{figure}
    \centering
    \includegraphics[width=1\linewidth]{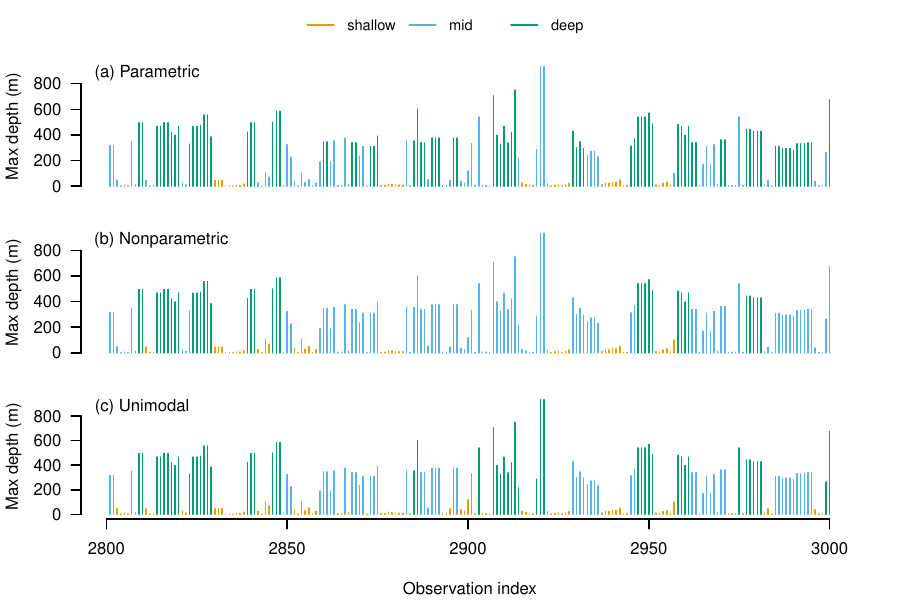}
    \caption{Segment of the time series coloured according to the decoded state sequence for the parametric model (a), the unstrained nonparametric model (b) and the nonparametric model with a unimodality constraint (c) fitted to the narwhal dive data.}
    \label{fig:narwhal_states}
\end{figure}

The estimated state-dependent distributions of the unconstrained nonparametric model look substantially different (Figure~\ref{fig:narwhal_models}, panel (c)). The increased flexibility enables the model to distinguish two deeper diving states, estimating a mode for state~2 around 350 metres. 
However, the narwhal tends to dive close to the surface for a few sampling units in between deeper dives (Figure \ref{fig:narwhal_states}) which is not adequately reflected by the Markovian state process (an issue that has been observed for marine mammals already by \citealp{borchers2013using}). 
When this is combined with the immense flexibility of P-spline-based state-dependent distributions, the model compensates for the rigidity of the state-process model by fitting a bimodal state-dependent distribution in the second state, increasing the likelihood of the associated state sequence under the Markov assumption. Hence, the emission distribution in state~2 had a similar behaviour to the second simulation experiment; it included the nonsensical mode at a depth of 0 metres.



Furthermore, the unconstrained model's flexibility in the tails of the distributions causes a problem when there is a short sequence of deep dives within a sequence of medium-depth dives: the deep dives will be classified in the same state as the medium-depth dives. 
As a result, state~2 is challenging to interpret as it comprises very small, mid-range, and very large observations  (Figure~\ref{fig:narwhal_models} (c) and (d)).



The nonparametric model with an added unimodality constraint provides a trade-off between a satisfactory fit to the empirical distribution of the dive-depth data, as well as a good separation between the three states (Figure ~\ref{fig:narwhal_models} (e) and (f)). This results in the three states being easily interpreted as \textit{shallow}, \textit{mid}, and \textit{deep} dives. Specifically, the distinction between dives of medium depth, and both shallow and deep dives are clearer than in other models (Figure~\ref{fig:narwhal_models}). Due to the unimodality constraints, the \textit{mid} dives state does not include near-surface dives, which are now attributed to the \textit{shallow} state (Figure~\ref{fig:narwhal_models} (e)), nor the extremely deep dives, which are now only attributed to the \textit{deep} state (Figure~\ref{fig:narwhal_models} (f)).

In Figure~\ref{fig:narwhal_states}, we see that the unimodal fit (panel (c)) classifies all dives with a small maximum depth as shallow. This is not the case for the parametric fit (panel (a)) and the unconstrained nonparametric fit (panel (b)). 
Due to these differences in state decoding, differences between the models also arise in the estimated number of state transitions. The parametric model estimates 305 behavioural switches while the unconstrained nonparametric model estimates only 217, likely because the Markov assumption favours high state persistence. The unconstrained model’s flexibility aggravates this tendency by allowing even longer-lasting stays. 
The unimodality constraint encourages distinct state patterns, thereby leading to a higher number (325) of estimated behavioural switches.


In summary, imposing an informative constraint --- reflecting prior knowledge that state-dependent densities are likely unimodal --- leads to a more stable and interpretable model fit while retaining most of the desirable flexibility of spline-based density estimation. It encourages more realistic state distinctions and prevents overlapping states, resulting from large state persistence.



\section{Discussion}


We have developed a simple method to enforce unimodality of estimates of state-dependent densities in HMMs. Our unimodal P-spline density estimates are extremely flexible, moving analysis beyond rigid parametric assumptions. This flexibility is controlled partly by the unimodality constraint, but also by a smoothness penalty, with automatic selection of the smoothing parameter weights. The estimation method is computationally efficient. To facilitate the use of constrained spline techniques in this and other modelling settings, we provide the relevant building blocks as easy-to-use functions in the \texttt{R} package \texttt{LaMa} \citep{koslikLaMa2025}.

Unimodality of a state-dependent density can make the state's meaning clear, and so unimodality is useful and often expected in real-world applications.
In addition, enforcing unimodality has certain robustness properties, as shown in our simulation studies, where we compare our method to the analogous P-spline method with no unimodality constraints.  
We find that little is lost by enforcing unimodality when simulated data are generated from an HMM with unimodal emission densities. But much can be gained when state transitions do not follow the usual HMM: our simulations show that fitting a P-spline-based HMM leads to better inference about the state sequence when unimodality is enforced. 
Thus, our method is, in some sense, robust to misspecification of the underlying state process, and so may overcome a common problem in HMM applications. While more flexible state process models can be considered, such as hidden semi-Markov models (HSMMs, see, for instance, \cite{suarez.2022.hsmm}), picking a suitable model might be difficult. In addition, the combination of HSMMs and nonparametric density estimation has not been developed, and would likely be computationally challenging.
Thus, unimodality-constrained nonparametric emission density estimation within an HMM offers a pragmatic analysis of time series data driven by a possibly complex state process.   

Our narwhal case study, which exhibited characteristics similar to one of our simulation scenarios, demonstrated a real-world situation in which using our method is advantageous. Neither a parametric HMM nor an unconstrained nonparametric HMM yielded a satisfactory balance between model fit and interpretability. The raw data indicated that there are likely three states, corresponding to shallow dives, mid-depth dives, and deep dives. 
The parametric HMM was not flexible enough to capture the three emission distributions. The unconstrained nonparametric HMM was too flexible, resulting in an uninterpretable state~2, which contained mid-depth dives and shallow dives, especially when shallow dives occured in short sequences between long sequences of mid-depth or deep dives. In addition, some deep dives were classified in state~2. These issues are most likely because the Markov assumption about the state sequence was not satisfied, so this misspecification forced the flexible nonparametric method to be even more flexible with the emissions distribution estimation.

Future research should explore strategies for further improving computational efficiency in automatic mode detection. In our current framework, a model should be  fit for each possible location of the mode in the coefficient sequence. However, the number of possible mode positions scales exponentially in the number of hidden states, making exhaustive exploration impractical.
We therefore proposed a two-step procedure to alleviate this burden, with the initial step using an unconstrained P-spline estimate to suggest the possible mode locations to consider. Developing a fully automatic and more efficient approach remains an important direction.
One approach would be to assume that each emissions density is log-concave, that is, that the logarithm of the density is a concave function, which ensures unimodality. While log-concave densities form a rich class of unimodal densities, they do not include unimodal densities with heavy tails. On the other hand, modelling a log-concave function with a form of B-splines is straightforward through inequality restrictions on the coefficients, and there is no need to specify a mode location a priori. However, ensuring that the resulting density is positive and integrates to one  may complicate the computation. \citet{wang_yan_2021} have developed an \texttt{R} package for constrained regression analysis, include concavity constraints, and this may be useful for estimating a log-concave density.
 
Unconstrained spline methods have been used to estimate transition probabilities, and a worthwhile extension would be the incorporation of shape constraints, particularly when transition probabilities include covariates. For instance, in analysing elk (\textit{Cervus elaphus}) tracking data, \citet{patterson2017statistical} modelled transition probabilities as depending parametrically on  distance to water. These could be modelled nonparametrically, using the approach of \citet{michelot2022hmmtmb}, who provides an \texttt{R} library 
for flexibly modelling not only the transition probabilities' dependence on a covariate but also the emissions densities. But one might place constraints on \citeauthor{patterson2017statistical}'s  model by, for instance, forcing a transition probability to be decreasing with distance from water. To study circadian rhythms, \citet{feldmann2023flexible} used time as a covariate, enforcing cyclic behaviour by estimating transition probabilities with penalised cyclic B-splines, instead of the usual sine and cosine functions \citep{huang_et_al.2018}. However, one might want to place shape restrictions on this dependence.  For example, an additional useful restriction on the shape of a cyclic transition probability might be to force the transition probability function to have only one maximum and one minimum per cycle. \citet{koslik2025tensor} models transition probability dependence on several covariates using tensor product splines, which allow for interactions among the covariates. \citeauthor{koslik2025tensor} applies the method to African elephant (\textit{Loxodonta}) data, where transition probabilities depend on time of day and day of year, with possible interactions between the two. Shape constraints on the cyclic dependence might be helpful here.

Another potential use of shape-restricted modelling is in Markov-switching regression models \citep{langrock2017markov}. Here, observations at time $t$ are a response variable $Y_t$ and a covariate vector ${\bf{X}}_t$. Of interest is the regression of $Y_t$ on ${\bf{X}}_t$, and the specific form of the regression depends on the underlying state. \citeauthor{langrock2017markov} have estimated these regression functions via unconstrained P-splines, but restrictions on the form of dependence might be worthwhile. For instance, in their application, the energy price ($Y_t$) depends on the oil price (${\bf{X}}_t$), but the form of the relationship depends on the economic regime, that is, the state of the economy. One might assume that, for each fixed economic regime, energy price increases with oil price.  
This idea could also be valuable in ecological applications. For example, \citet{byrnes2023daily} modelled bull shark (\textit{Carcharhinus leucas}) activity using overall dynamic body acceleration as a proxy for movement intensity. 
As an ectothermic species, bull shark activity levels strongly depend on ambient temperature. Hence, the mean activity level could be modelled explicitly as a function of temperature. In such a context, it may be biologically reasonable to impose shape restrictions on these functions --- such as monotonicity or unimodality --- to reflect the expectation that activity increases with temperature up to an optimal range before declining.

Both Markov switching models and the incorporation of covariates into transition probabilities are related to regression, and so are free of the cumbersome restrictions of density estimation. Much literature exists in the constrained regression setting, and this work might be useful for further work in HMMs. \citet{pya2015shape} have developed  methodology for constrained nonparametric regression analysis in a non-HMM context using generalized additive models. \citet{Wang.Taylor.2004} and \citet{bollaerts.2006} develop flexible multiple regression methods by considering  shape constrained tensor-product splines.
Tensor product splines, without constraints, have been used in HMMs by \citet{michels2025nonparametric}  to estimate bivariate emissions distributions.

Overall, our approach offers a pragmatic balance between flexibility, robustness, and interpretability. These characteristics make the approach especially valuable for practitioners, providing models that are both reliable and easy to apply in real-world settings.




\section*{Supplementary Materials}
The Supplementary Materials include the data and code required to reproduce all simulation experiments and the narwhal case study, as well as additional figures. An in-depth tutorial is also provided, demonstrating how models with B-spline-based densities and unimodality constraints can be implemented using the \texttt{R} package \texttt{LaMa}. 
All data, code, and the tutorial are available from the public GitHub repository: \url{https://github.com/janoleko/unimodalHMMs}.

\section*{Acknowledgments}

The authors thank Roland Langrock for helpful comments on an earlier version of this manuscript. 
Marie Auger-Méthé thanks the Natural Sciences and Engineering Research Council of Canada (NSERC) Discovery Grant and Northern Supplement programs, the Canada Research Chairs program, BC Knowledge Development fund, Canada Foundation for Innovation’s John R. Evans Leaders Fund, and the Canadian Statistical Sciences Institute (CANSSI).  
Nancy Heckman thanks the Natural Sciences and Engineering Research Council of Canada (NSERC) Discovery Grant program.
We are thankful to the community of Mittimatalik and the Mittimatalik Hunter and Trapper Organization for supporting our narwhal tagging projects. A special thanks to the team that assisted with capturing and handling narwhals in the field. Field work was supported by the Polar Continental Shelf Program, Fisheries and Oceans Canada, the Nunavut Wildlife Management Board, Nunavut Implementation Fund, World Wildlife Fund Canada, and the University of Windsor.

\printbibliography

\newpage

\appendix
\renewcommand{\thesubsection}{\Alph{section}.\arabic{subsection}}

\section{Appendix}

\label{A1: additional figures}

\begin{figure}[H]
    \centering
    \includegraphics[width=1\linewidth]{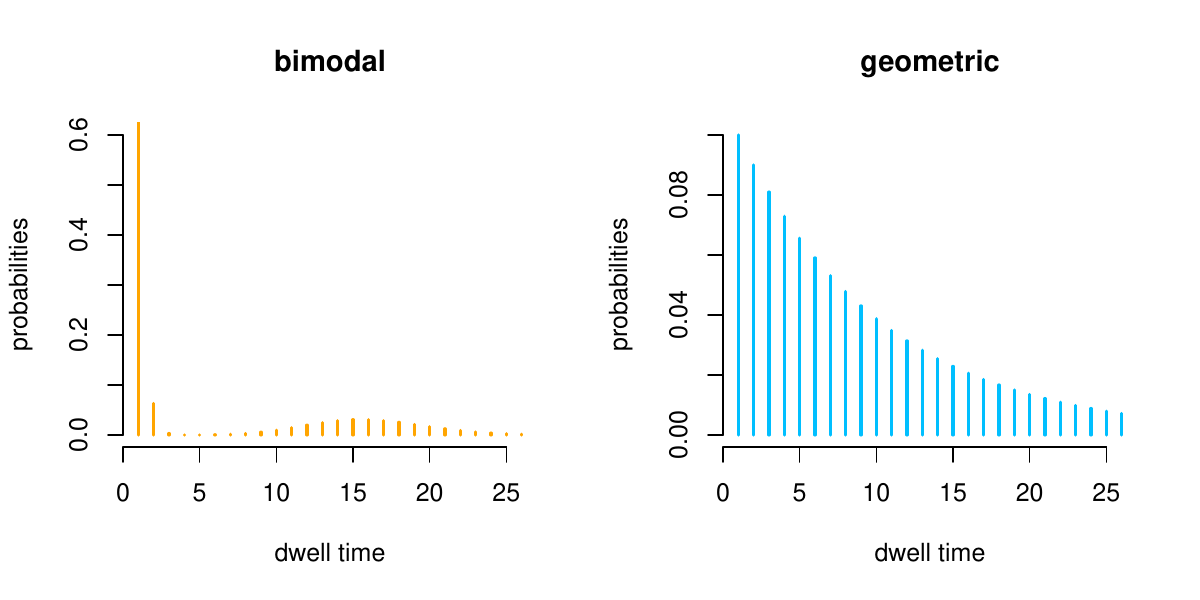}
    \caption{State dwell time distributions of the hidden semi-Markov model used to simulate the data for the second simulation experiment. Bimodal probability mass function for state~1 (left) and geometric probability mass function for state~2 (right).}
    \label{fig:hsmm_sim_dwelltime}
\end{figure}






\end{document}